\documentclass[12pt]{iopart}
\usepackage{graphicx}
\usepackage{comment}
\usepackage{iopams}
\usepackage{revsymb}
\begin{document}
\title{Cavity-Assisted Back Action Cooling of Mechanical Resonators}

\author{I~Wilson-Rae$^{1,*}$, N~Nooshi$^1$, J~Dobrindt$^2$,
  TJ~Kippenberg$^2$, and W~Zwerger$^1$} 
\address{$^1$Technische Universit\"at M\"unchen, D-85748 Garching,
  Germany}  
\address{$^2$Max Planck Institut f\"{u}r Quantenoptik, D-85748
  Garching, Germany.} 
\ead{${}^{*}$Ignacio.Wilson-Rae@ph.tum.de}

\date{\today}

\begin{abstract}
  We analyze the quantum regime of the dynamical backaction cooling of
  a mechanical resonator assisted by a driven harmonic oscillator
  (cavity). Our treatment applies to both optomechanical and
  electromechanical realizations and includes the effect of thermal
  noise in the driven oscillator. In the perturbative case, we derive
  the corresponding motional master equation using the
  Nakajima-Zwanzig formalism and calculate the corresponding output
  spectrum for the optomechanical case. Then we analyze the strong
  optomechanical coupling regime in the limit of small cavity
  linewidth. Finally we consider the steady state covariance matrix of
  the two coupled oscillators for arbitrary input power and obtain an 
  analytical expression for the final mechanical occupancy.  This is
  used to optimize the drive's detuning and input power for an
  experimentally relevant range of parameters that includes the
  ``ground state cooling'' regime.
\end{abstract}


\maketitle 

\section{Introduction}

Recent progress in the emerging fields of nanoelectromechanical
\cite{Craighead00,Ekinci05} and optomechanical systems
\cite{Braginsky77,Kippenberg07} promises to enable quantum limited
control of a single macroscopic mechanical degree of freedom
\cite{LaHaye04,Knobel03}.  This is relevant in the context of high
precision measurements \cite{Caves81}-
\cite{Bocko96}
like single spin magnetic resonance force microscopy
\cite{Sidles95,Gassmann04} (MRFM) and for fundamental studies of the
quantum to classical transition \cite{Schwab05,Vitali07b,Vitali07a}.
A paradigmatic goal which has triggered a surge of activity is to
prepare the eingenmode associated to an ultra long lived mechanical
resonance (angular frequency $\omega_m$ and $Q$-value $Q_m$) in its
quantum ground state with high fidelity
\cite{Mancini98}-%
\cite{Teufel08}.
The considerable difficulty to achieve the desired combination $k_B T
\ll \hbar \omega_m$, $Q_m\gg1$ with state of the art micro-fabrication
and cryogenic techniques \cite{Ekinci05} has naturally motivated ideas
to use cooling schemes analogous to the laser-cooling of atoms
\cite{Wineland79}-
\cite{Ashkin97}. 

\begin{figure}
\centerline{\includegraphics[width=.6\textwidth]{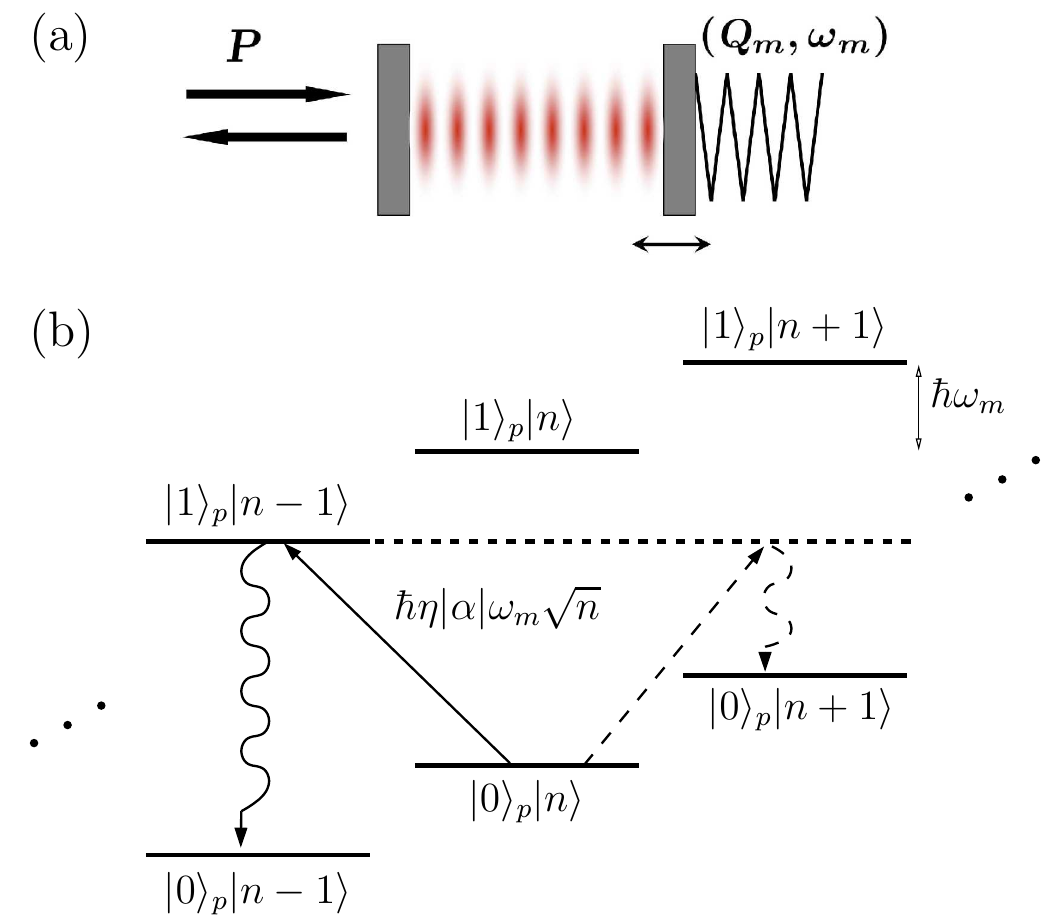}}
\caption{(a) Fabry-P\'{e}rot equivalent of a mechanical eigenmode
  (frequency $\omega_m/2\pi$ and $Q$-value $Q_m$) coupled to an
  optical cavity mode. (b) Level diagram of the two modes in a
  ``shifted'' representation for perturbative optomechanical coupling
  $\eta$. Raman scattering processes induced by the latter can
  decrease (solid lines) or increase (dashed) the mechanical
  eigenmode's quantum number $n$ ($\alpha$ is the steady state
  amplitude in the cavity mode and $|0\rangle_p,\,|1\rangle_p,\ldots$
  its Fock states).
 [Reproduced from I.~Wilson-Rae, N.~Nooshi, W.~Zwerger, and
 T.J.~Kippenberg. Theory of ground state cooling of a mechanical
 oscillator using dynamical backaction. Phys. Rev. Lett., 99:093901,
 2007. ``Copyright (2007) by the American Physical Society.'']
  \label{fig:FP}}
\end{figure}

In these schemes the mechanical resonator's displacement is coupled
parametrically to an auxiliary high frequency bosonic or fermionic
resonator (pseudospin) that can act as a ``cooler''
\cite{Braginsky02}. To drive the latter while monitoring its output
allows detection of the mechanical displacement. Naturally there is a
back-action force associated to this measurement process
\cite{Caves80,Braginsky92,Braginsky96}. Due to the dissipative
dynamics of the cooler for a negative detuning of the drive this force
becomes anti-correlated with the Brownian motion resulting in net
cooling. In turn the quantum fluctuations of the cooler --- which in
the atomic laser cooling manifest in the inherent stochastic nature of
the spontaneous photon emissions --- result in a quantum noise
spectrum for this backaction force that sets a fundamental lower bound
for the final temperature \cite{Stenholm86,Leibfried03}. Thus the
structured reservoir afforded by the driven cooler and its environment
provides an effective thermal bath for the mechanical resonator. The
concomitant absorptions of motional quanta (cooling) correspond to
Raman scattering processes in which a drive quanta is up converted,
while emission events (heating) are associated to Raman processes in
which a drive quanta is down converted [cf.~Fig.~\ref{fig:FP}(b)].

A host of concrete realizations of the above generic scenario have
been discussed in the literature. These range from electronic or
electrical devices where the cooler is provided by a (superconducting)
single electron transistor \cite{Naik06}, a Cooper-pair box
\cite{Martin04}, an LC-circuit \cite{Braginsky77}, or a quantum dot
\cite{WilsonRae04}; to optomechanical systems where this role is
played by an optical cavity mode \cite{Kippenberg07}. In the latter
systems, which are equivalent to a Fabry-Perot with a moving mirror
[cf.~Fig.~\ref{fig:FP}(a)], the optical field couples parametrically
to the mechanical motion via radiation pressure. This effect has
already been thoroughly demonstrated experimentally
\cite{Kippenberg05} and harnessed to provide appreciable cooling,
albeit in the classical regime
\cite{Gigan06,Arcizet06,Schliesser06,Thompson08,Schliesser08}.
Recent experiments involve both optical and microwave cavities. In
turn completely analogous effects have been shown using a capacitively
coupled radio frequency LC-circuit \cite{Brown07} and a superconducting
microwave cavity \cite{Teufel08}. These systems share
the common feature that the cooler is well approximated by a single
harmonic oscillator which henceforth will be referred to as the
``cavity''. While for optical cavities the vacuum constitutes an
excellent approximation for the input when the drive is switched off,
in the case of radio and microwave frequencies thermal noise in the
cavity needs to be taken into account. A quantum treatment of the
corresponding temperature limits has already been given which predicts
that ground state cooling is possible when the mechanical oscillation
frequency is larger than the cavity's linewidth $\kappa$ 
\cite{WilsonRae07,Marquardt07,Bhattacharya07}. Here we provide a
rigorous analysis of the cooling dynamics that drives the mechanical
resonator mode to a thermal state with a well defined effective final
temperature that for a finite $Q_m$ is imprinted on the cavity's
output. This is done in Sec.~\ref{pert} where we obtain the motional
master equation and the corresponding output spectrum.

The above picture is only valid for perturbative optomechanical
coupling and thus breaks down when the resulting cooling rate becomes
comparable to the cavity linewidth $\kappa$ or to the mechanical
oscillation frequency \cite{Marquardt07} $\omega_m$. In the Doppler
limit $\omega_m\ll\kappa$ the system then approaches the parametric
instability or settles into a regime where backaction effects are
purely diffusive with no net cooling. On the other hand in the
resolved sideband limit $\kappa\ll\omega_m$ as the optomechanical
coupling exceeds the cavity linewidth the system enters into a strong
coupling regime in which the motion hybridizes with the cavity
fluctuations. The ensuing optomechanical normal modes are then cooled
simultaneously. This phenomenon is analyzed in Sec.~\ref{nonpert} in
the limit of small cavity linewidth where we find that the dynamics of
each of the normal modes can be described by a master equation
analogous to the one valid in the perturbative regime with a cooling
rate given by half the cavity linewidth. Finally, in the same Section
we derive an analytical expression for the final (steady state)
average mechanical occupancy (phonon number) valid for arbitrary
optomechanical coupling and use it to optimize the parameters of the
drive.

\section{Optomechanical Master Equation}

In optomechanical systems (and their electromechanical analogues) the
cavity resonant frequency $\omega_p$ depends inversely on a
characteristic length that is modified by the mechanical resonator's
displacement. The fact that $\omega_p\gg\omega_m$ allows for an
adiabatic treatment of this effect which results in the aforementioned
parametric coupling. In general the leading contribution to the
latter is linear in the mechanical displacement --- but situations in
which it is instead quadratic can be readily engineered
\cite{Bhattacharya07,Thompson08}. The desired cooling dynamics is
induced by a slightly detuned electromagnetic drive (angular frequency
$\omega_L$), which for optomechanical systems corresponds to an
incident laser and in the electromechanical case is afforded by a
suitable external AC voltage. Thus the Hamiltonian describing the
coupled system (in a rotating frame at $\omega_L$) is given by
\cite{Law95}
\begin{equation}\label{eq:hamun}
H'= -\hbar\Delta_L'a_p^{\dag}a_p + \hbar\eta\omega_m
a_p^{\dag}a_p(a_m+a_m^{\dag})+\hbar\frac{\Omega}{2}(a_p+a_p^{\dag})+
\hbar\omega_m a_m^{\dag}a_m \,.
\end{equation}
Here $a_p$ ($a_m$) is the annihilation operator for the
electromagnetic (mechanical) oscillator and $\Delta_L'$ is the
detuning of the drive from $\omega_p$. Here we have defined $\Omega
\equiv 2\sqrt{P\kappa_\mathrm{ex}/\hbar\omega_L}$, where $P$ is the
input power of the drive and $\kappa_\mathrm{ex}$ is the cavity decay
rate into the associated outgoing electromagnetic modes. The
dimensionless parameter $\eta$, characterizing the non-linear coupling
between the cavity and the mechanical resonator, is given by
$\eta=(\omega_p/\omega_m)(x_0/L)$ where $x_0$ is the zero point motion
of the mechanical resonator mode and the characteristic length $L$
depends on the physical realization. In the optomechanical case it
corresponds to an effective optical cavity length while for
electromechanical realizations \cite{Brown07} $L=2d
C_\mathrm{tot}/C_c$, where $C_c\propto 1/d$ is the dynamical
capacitance, $d$ is the distance between the corresponding electrodes,
and $C_\mathrm{tot}$ is the total capacitance.

We treat the losses induced by the electromagnetic and mechanical
baths within the rotating wave Born-Markov approximation using the
standard Lindblad form Liouvillians \cite{Gardiner}. It is important
to note that the validity of a rotating wave approximation (RWA) in
the environmental coupling responsible for the mechanical losses only
amounts to $Q_m\gg1$ provided the optomechanical coupling is weak
enough that there is no appreciable mixing between the annihilation
and creation operators of the modes (cf.~Sec.~\ref{nonpert}). Clearly
if the latter is not satisfied the usual RWA will result in the
unwarranted neglection of resonant terms. This can be borne out by
comparing the corresponding displacement spectra and results in the
condition
$\eta|\alpha|\omega_m\ll\max\{\sqrt{\omega_m\kappa},\,\omega_m\}$. Henceforth
we focus on parameters that satisfy it which include the most relevant
regimes for cooling and ensure that the system doe not approach the
instability for any detuning. Thus the evolution for the density
matrix of the resonator-cavity system reads
\begin{eqnarray}\label{eq:masteropt}
  \fl \dot{\rho}=-\frac{i}{\hbar}[H',\rho]+\frac{\kappa}{2}n(\omega_p)
  (2a_p^{\dag}\rho
  a_p-a_pa_p^{\dag}\rho-\rho a_pa_p^{\dag}) \nonumber \\
  +\frac{\kappa}{2}[n(\omega_p)+1](2a_p\rho
  a_p^{\dag}-a_p^{\dag}a_p\rho-\rho a_p^{\dag}a_p)  \nonumber \\
  +\frac{\gamma_m}{2}n(\omega_m)(2a_m^{\dag}\rho
  a_m-a_ma_m^{\dag}\rho-\rho a_ma_m^{\dag}) \nonumber \\
  +\frac{\gamma_m}{2}[n(\omega_m)+1](2a_m\rho
  a_m^{\dag}-a_m^{\dag}a_m\rho-\rho a_m^{\dag}a_m)\,.
\end{eqnarray}
The total cavity decay rate $\kappa$ 
has two contributions: (i) the rate at which photons are lost from the
open port (where the driving field comes in) given by
$\kappa_\mathrm{ex}$ and (ii) the ``internal loss'' rate $\kappa
-\kappa_\mathrm{ex}$ due to the other losses of the electromagnetic
resonator (i.e absorption inside the dielectric, scattering into other
modes, etc.). Naturally the thermal noise is determined by the Bose
number $n(\omega)$. At room temperature and for optical frequencies
$n(\omega_p)$ is negligible, however for the much lower radio and
microwave frequencies characterizing electromechanical setups this
quantity can be comparable to the final mechanical occupancies
achieved. Similarly $\gamma_m=\omega_m/Q_m$ is the mechanical
resonator's natural linewidth and $n(\omega_m)$  its mean
occupation number at thermal equilibrium (i.e.~in the absence of the
drive).


To study the cooling process it proves useful to apply a canonical
transformation of the form $a_p\to a_p+\alpha$, $a_m\to a_m+\beta$
with the amplitudes $\alpha$, $\beta$ chosen so that the linear terms
in the transformed Liouvillian cancel out. This condition leads to the
following coupled equations for the amplitudes
\begin{eqnarray}
\Omega -(\Delta_L'+i\frac{\kappa}{2}) \alpha+ \eta \omega_m \alpha
(\beta+\beta^*) = 0 \,, \nonumber \\ 
(\omega_m-i\frac{\gamma_m}{2})\beta+\eta\omega_m|\alpha|^2=0 \,.
\end{eqnarray}
We assume $\eta\ll1$ and that the mechanical dissipation rate
$\gamma_m$ is much smaller than $\omega_m$. To lowest order in the
small parameters $\eta$ and $1/Q_m$ we obtain
$\alpha\approx\Omega/(2\Delta_L'+i\kappa)$ and
$\beta\approx-\eta|\alpha|^2$. Here $|\alpha|^2$ is the steady state
occupancy of the cavity in the absence of optomechanical coupling and
$\beta$ is the static shift of the mechanical amplitude due to the
radiation pressure.  The normal coordinates after the transformation
are shifted so that the new amplitudes correspond to the deviation
from the steady state equilibrium position. This transformation leaves
the dissipative part of the Liouvillian invariant and transforms the
Hamiltonian into:
\begin{equation}\label{eq:hams}
  H=-\hbar\Delta_La_p^{\dag}a_p+\hbar\omega_ma_m^{\dag}a_m+
  \hbar\eta\omega_m (a_p^{\dag}a_p+\alpha^*a_p+\alpha
  a_p^{\dag})(a_m+a_m^{\dag})\,, 
\end{equation}
where we have introduced the effective detuning
$\Delta_L'+2\eta^2|\alpha|^2\omega_m\rightarrow\Delta_L$. It is
interesting to note that if the bosonic cooler is replaced by a
fermionic one so that $a_p\to\sigma_-$ we obtain a Hamiltonian that
resembles the one describing a trapped ion in the Lamb-Dicke
regime. As a result for perturbative $\eta$ (as analyzed in the next
Section) the cooling cycle [cf.~Fig.~\ref{fig:FP}(b)] becomes
analogous to the Lamb-Dicke regime of atomic laser-cooling
\cite{WilsonRae07}.

\section{Perturbative Cooling}\label{pert}

\subsection{Master equation for mechanical motion}\label{master}

We first focus on the regime in which the input laser power $P$ is low
enough so that the time scales over which the populations of the
mechanical resonator's Fock states evolve (leading to cooling or
heating) are much slower than those associated to the losses of the
cavity and to the free mechanical frequency. As will become clear
below this requires $\eta^2 |\alpha|^2 \ll (k/\omega_m)^2$.

Here we also assume $[n(\omega_m) + 1] \gamma_m \ll \kappa$,
$\omega_m$ which must hold to allow for appreciable cooling, and
$\eta^2\ll 1$ (in optomechanical realizations $\eta \lesssim
10^{-4}$). Hence the electromagnetic degrees of freedom can be treated
as a structured environment that affects the mechanical motion
perturbatively. Along these lines the latter can be described by a
Markovian master equation for its reduced density matrix
\cite{Gardiner}. 

To derive it we take the optomechanical master equation in the shifted
representation, transform to an interaction picture for the resonator
mode and adiabatically eliminate the cavity using the Nakajima-Zwanzig
formalism \cite{Zwanzig64,Gardiner,Cirac92}.  The optomechanical
coupling and the mechanical losses are treated perturbatively. We
define the projection
\begin{equation}\label{eq:projection}
\mathcal{P} \rho = \Tr_p \{\rho\} \otimes \rho_p^{(\mathrm{th})}
\qquad\quad \mathcal{Q} \equiv \openone - \mathcal{P} \,,
\end{equation}
with
\begin{equation}\label{eq:rhoth}
\rho_p^{(\mathrm{th})}=\frac{1}{n_p+1} \sum_{n=0}^\infty
\left[ \frac{n_p}{n_p+1} \right]^n | n \rangle \langle
n |_p \,,
\end{equation}
where $n_p\equiv n(\omega_p)$ is the Bose number, and introduce the
formal parameter $\zeta$ such that
\begin{equation}\label{eq:L}
  \mathcal{L}(t) = \zeta^2 \mathcal{L}_0 +
  \zeta\mathcal{L}_1(\zeta^2t) + \mathcal{L}_2 \,,
\end{equation}
with 
\begin{eqnarray}
  \fl \mathcal{L}_0 \rho \equiv i [\Delta_L a_p^\dag
  a_p^{\vphantom{\dag}}, \rho ] + \frac{\kappa}{2} n (\omega_p) [2
  a_p^\dag \rho a_p^{\vphantom{\dag}} - a_p^{\vphantom{\dag}} a_p^\dag
  \rho - \rho a_p^{\vphantom{\dag}} a_p^\dag ] \nonumber\\ +
  \frac{\kappa}{2} [n (\omega_p)+1] [ 2 a_p^{\vphantom{\dag}} \rho
  a_p^\dag - a_p^\dag a_p^{\vphantom{\dag}} \rho - \rho a_p^\dag
  a_p^{\vphantom{\dag}} ] \,,\label{eq:L0}\\ \fl
  \mathcal{L}_1(\zeta^2t) \equiv
  e^{i \omega_m \zeta^2 t} \mathcal{L}_1^{(+)} + e^{-i \omega_m
  \zeta^2 t} \mathcal{L}_1^{(-)} \,,\label{eq:L1}\\ \fl
  \mathcal{L}_1^{(+)} \rho \equiv -i \left[\eta \omega_m \left[
  a_p^\dag a_p^{\vphantom{\dag}} +
  \alpha^*a_p^{\vphantom{\dag}}+\alpha a_p^{\dag} \right] a_m^\dag,
  \rho \right] \,,\label{eq:L1p} \\ \fl \mathcal{L}_1^{(-)} \rho \equiv -i
  \left[\eta \omega_m \left[ a_p^\dag a_p^{\vphantom{\dag}} +
  \alpha^*a_p^{\vphantom{\dag}}+\alpha a_p^{\dag} \right]
  a_m^{\vphantom{\dag}}, \rho \right] \,,\label{eq:L1m} \\ \fl
  \mathcal{L}_2 \equiv \frac{\gamma_m}{2} n(\omega_m) (2 a_m^\dag \rho
  a_m^{\vphantom{\dag}} - a_m^{\vphantom{\dag}} a_m^\dag \rho - \rho
  a_m^{\vphantom{\dag}} a_m^\dag ) \nonumber\\ + \frac{\gamma_m}{2}
  [n(\omega_m)+1] ( 2 a_m^{\vphantom{\dag}} \rho a_m^\dag - a_m^\dag
  a_m^{\vphantom{\dag}} \rho - \rho a_m^\dag
  a_m^{\vphantom{\dag}})\,.\label{eq:L2}
\end{eqnarray}
We note that $\mathcal{P}\rho$ is a stationary state of $\mathcal{L}_0$
for any $\rho$ implying that $\mathcal{L}_0\mathcal{P}=0$ while
$\mathcal{P}\mathcal{L}_0=0$ follows from trace preservation, so that
we have
\begin{equation}\label{eq:proj}
  \mathcal{Q}\mathcal{L}_0\mathcal{Q}=\mathcal{L}_0\,, \qquad\quad
  \mathcal{P}\mathcal{L}_0\mathcal{P}=\mathcal{Q}\mathcal{L}_0\mathcal{P}
  =\mathcal{P}\mathcal{L}_0\mathcal{Q}=0\,. 
\end{equation}
As will emerge from our derivation the basic idea is that the rates
for cooling and heating set the relevant time scale (zeroth order in
$1/\zeta$) which is widely separated from the mechanical oscillation
period $2\pi/\omega_m$ and the cavity ring down $1/\kappa$ (order
$1/\zeta^2$). In fact, the asymptotic expansion for $\zeta\to\infty$
pursued below amounts to a controlled expansion in the ratio between
the fast and the slow timescales. Given that we are interested in the
behaviour as $t\to\infty$, the initial condition is immaterial and we
choose for simplicity one in the $\mathcal{P}$-manifold so that
$\mathcal{Q}\rho_0=0$. Subsequently by explicitly integrating the
differential equation for the irrelevant part $\mathcal{Q} \rho$, we
obtain a closed equation for the relevant part $\mathcal{P} \rho$:
\begin{equation}\label{eq:mastex}
  \fl \mathcal{P} \dot{\rho} = \mathcal{P} \mathcal{L}(t) \mathcal{P}
  \rho + \mathcal{P} \mathcal{L}(t) \int_0^t \!\rmd \tau\,
  \mathcal{T}_+\! \left[ e^{\int_0^t \!\rmd \tau' \mathcal{Q}
  \mathcal{L}(\tau') \mathcal{Q}} \right] \mathcal{T}_-\! \left[ e^{-
  \int_0^\tau \!\rmd \tau'' \mathcal{Q} \mathcal{L}(\tau'')
  \mathcal{Q} } \right] 
  \mathcal{L}(\tau)\mathcal{P} \rho(\tau) \,,
\end{equation}
where $\mathcal{T}_+$ ($\mathcal{T}_-$) is the time-ordering
(anti-time-ordering) operator.
For the purpose of analyzing the asymptotic limit $\zeta\to\infty$ of
Eq.~(\ref{eq:mastex}) we have 
\begin{equation}\label{eq:approxS}
  \fl  \mathcal{T}_+ \left[ e^{\int_0^t \!\rmd \tau' \mathcal{Q}
      \mathcal{L}(\tau') \mathcal{Q}} 
  \right] \mathcal{T}_- \left[e^{- \int_0^\tau \!\rmd \tau''
      \mathcal{Q} \mathcal{L}(\tau'') \mathcal{Q}} \right] =
  e^{\zeta^2  \mathcal{Q} \mathcal{L}_0 \mathcal{Q} (t-\tau)} \left[
    \openone + \Or\!\left( \frac{1}{\zeta} \right) \right]  \,,
\end{equation}
which can be understood considering the corresponding Laplace
transforms. Substitution of Eqs.~(\ref{eq:proj}), (\ref{eq:approxS})
into Eq.~(\ref{eq:mastex}) and the change of variables
$\tau'=\zeta^2(t-\tau)$ then yields
\begin{eqnarray}\label{eq:mastapprox1}
  \fl \mathcal{P} \dot{\rho} =
  \mathcal{P}\left[\zeta\mathcal{L}_1(\zeta^2t) + \mathcal{L}_2
  \right]\mathcal{P}\rho\nonumber\\ +\,
  \mathcal{P}\mathcal{L}_1(\zeta^2t)\mathcal{Q} \int_0^{\zeta^2 t}
  \!\rmd \tau'\, e^{\mathcal{L}_0 \tau'}
  \mathcal{Q}\mathcal{L}_1(\zeta^2t-\tau')\mathcal{P}
  \rho(t-\tau'/\zeta^2) + \ldots
\end{eqnarray}
Here we have also used $\mathcal{Q}^2=\mathcal{Q}$,
$\mathcal{P}^2=\mathcal{P}$, and
$\mathcal{P}\mathcal{L}_2=\mathcal{L}_2\mathcal{P}$. The leading order
in $1/\zeta$ of the evolution over the aforementioned relevant time
scale will be determined by the limit as $\zeta\to\infty$ of the
Laplace transform of the above. In the time domain all the fast
rotating terms (frequencies of order $\zeta^2 \omega_m$) drop out and
Eq.~(\ref{eq:mastapprox1}) reduces to
\begin{equation}\label{eq:mastapprox2}
  \fl \mathcal{P} \dot{\rho} =
  \mathcal{P}\mathcal{L}_2\mathcal{P}\rho
  \,+\, \left[\mathcal{P}\mathcal{L}_1^{(+)}\mathcal{Q}
    \int_0^{\infty} \!\rmd
    \tau'\, e^{(i\omega_m+\mathcal{L}_0) \tau'}\mathcal{Q}
    \mathcal{L}_1^{(-)}\mathcal{P}\rho + \mathrm{H.c.}\right]\,. 
\end{equation}
We note that as $\mathcal{P} =\lim_{t\to\infty} e^{\mathcal{L}_0t}$ we have
\begin{equation}\label{eq:eigenQ}
 \mathcal{L}_0 \rho=0 \qquad \Rightarrow \qquad \mathcal{Q}\rho=0 \,.
\end{equation}
It follows that the restriction of $\mathcal{L}_0$ to the
$\mathcal{Q}$-manifold has only eigenvalues with negative real
parts which allows us to establish
\begin{equation}\label{eq:intL0}
   \mathcal{Q} \int_0^{\infty} \!\rmd
  \tau'\, e^{(i\omega_m+\mathcal{L}_0) \tau'}\mathcal{Q} =
  \mathcal{Q}\left(-i\omega_m-\mathcal{L}_0\right)^{-1}\mathcal{Q}\,. 
\end{equation}

We focus on the parameter regime where $|\alpha|^2\gg n_p$. The
behavior of the cavity correlations that determine the second term in
Eq.~(\ref{eq:mastapprox2}) implies that in this regime contributions
arising from the cubic term in $\mathcal{L}_1^{(\pm)}$ are negligible
compared to those generated by the quadratic term --- for $n_p=0$ the
contributions of the former are higher order in $\eta$. The conditions
that warrant this linearization of $\mathcal{L}_1^{(\pm)}$ for the
case $n_p=0$ will be discussed further in the next Subsection. The
restriction of the quadratic term to the $\mathcal{P}$-manifold
vanishes and we obtain
\begin{eqnarray}\label{eq:intL0ap}
  \fl\Tr_p \left\{\mathcal{P}\mathcal{L}_1^{(+)}\mathcal{Q}
    \int_0^{\infty} \!\rmd \tau'\, e^{(i\omega_m+\mathcal{L}_0)
    \tau'}\mathcal{Q} \mathcal{L}_1^{(-)}\mathcal{P}\rho\right\}
  \nonumber\\
  \approx - 
\frac{g_m^2}{2} \left\{G(\omega_m,n_p)\left[
    a_m^{\dag},a_m^{\vphantom{\dag}}\mu\right] -G^{*}(-\omega_m,n_p) \left[
    a_m^{\dag},\mu a_m^{\vphantom{\dag}}\right]\right\} \,.
\end{eqnarray}
Here we have used Eqs.~(\ref{eq:projection}),
(\ref{eq:L1p}), (\ref{eq:L1m}), and introduced the optomechanical
coupling
\begin{equation}\label{eq:gm}
g_m\equiv 2 \eta |\alpha| \omega_m \,,
\end{equation}
the reduced density matrix for the mechanical mode
$\mu\equiv\Tr_p\{\rho\}$, and the cavity-quadratures' correlations
\begin{equation}\label{eq:corr1}
G(\omega,n_p) = \int_0^{\infty} \!\rmd \tau\,
  e^{i\omega \tau} \Tr_p
  \left\{X_p(0)e^{\mathcal{L}_0\tau}X_p(0)\rho_p^{(\mathrm{th})}\right\}\,,
\end{equation}
with $X_p(0)\equiv (\alpha^*a_p^{\vphantom{\dag}}+\alpha
a_p^{\dag})/\sqrt{2}|\alpha|$.  If we substitute
Eqs.~(\ref{eq:projection}), (\ref{eq:L2}), (\ref{eq:intL0ap}) into
(\ref{eq:mastapprox2}), trace out the cavity, and rearrange we finally
obtain the following master equation
\begin{eqnarray}\label{eq:master}
\fl  \dot{\mu}=-i[(\omega_m+\Delta_m)a_m^{\dag}a_m,\mu] \nonumber\\ 
+ \frac{1}{2}\lbrace\gamma_m[n(\omega_m)+1]+A_{-}(n_p) \rbrace
(2a_m\mu a_m^{\dag}-a_m^{\dag}a_m\mu-\mu a_m^{\dag}a_m) \nonumber\\ 
+\frac{1}{2}[\gamma_mn(\omega_m)+A_{+}(n_p)](2a_m^{\dag}\mu a_m-
a_ma_m^{\dag}\mu-\mu a_ma_m^{\dag})\,,
\end{eqnarray}
where the cooling (heating) rate $A_{-}(n_p)$ [$A_{+}(n_p)$] and the
mechanical frequency shift $\Delta_m$ induced by the optomechanical
coupling are given by
\begin{eqnarray}\label{eq:A1}
A_{\mp}(n_p)=g_m^2 \Re \{ G(\pm \omega_m, n_p) \} \,,\\ \Delta_m =
\frac{g_m^2}{2} \Im \{ G(\omega_m, n_p) - G(-\omega_m, n_p)\} \,.\label{eq:D1}
\end{eqnarray}
It is straightforward to calculate the necessary two-time correlations
using the quantum regression theorem given the steady state moments:
\begin{equation}\label{eq:moments}
\langle a_p^{\vphantom{\dag}}a_p^{\dag}\rangle= n_p+1 \,, \qquad \langle
a_p^{\dag} a_p^{\vphantom{\dag}}\rangle= n_p \,, \qquad \langle
a_p^{\vphantom{\dag}} a_p^{\vphantom{\dag}}\rangle= 0 \,,
\end{equation}
and that the evolution of the mean amplitude reads
\begin{equation}\label{eq:amp}
\langle \dot{a}_p\rangle= \left( i \Delta_L - \kappa/2 \right) \langle
{a}_p\rangle \,.
\end{equation}
Thus we obtain
\begin{equation}\label{eq:corr2}
G(\omega, n_p) = \left[ G (\omega, 0) + G^* (\omega, 0) \right] n_p +
G (\omega, 0)  \,,
\end{equation}
with 
\begin{equation}\label{eq:corr0}
G(\omega,0) = \frac{1}{-2i (\omega+\Delta_L) + \kappa}
\end{equation}
which substituted into Eqs.~(\ref{eq:A1}), (\ref{eq:D1}) leads to
\begin{eqnarray}\label{eq:A2}
  A_{\mp}(n_p)=[A_{-}(0)+A_{+}(0)]n_p+A_{\mp}(0) \,, \\ \Delta_m
  = 
  g_m^2 \left[\frac{\Delta_L- \omega_m}{4
      (\Delta_L-\omega_m)^2 +\kappa^2}+ \frac{\Delta_L+ \omega_m}{4(
      \Delta_L+\omega_m)^2+\kappa^2}\right] \,,\label{eq:D2}
\end{eqnarray}
where the corresponding rates in the absence of thermal noise in
the driven cavity are given by:
\begin{equation}\label{eq:A0}
 A_{\mp}(0)= g_m^2
 \frac{\kappa}{4(\Delta_L\pm\omega_m)^2+\kappa^2}\,.
\end{equation}
The above can be related to the input power and the frequency of the
drive via:
\begin{eqnarray}
  g_m = 2 \eta \omega_m \sqrt{\frac{P \kappa_{ex}}{\hbar \omega_L 
      (\Delta_L^2 + \kappa^2/4)}} \,, \\
  \Delta_L = \omega_L-\omega_p +2\eta^2\omega_m\frac{P
    \kappa_{ex}}{\hbar \omega_L (\Delta_L^2 + \kappa^2/4)}\,. 
\end{eqnarray}

\begin{figure}
\centerline{\includegraphics[width=.4\linewidth]{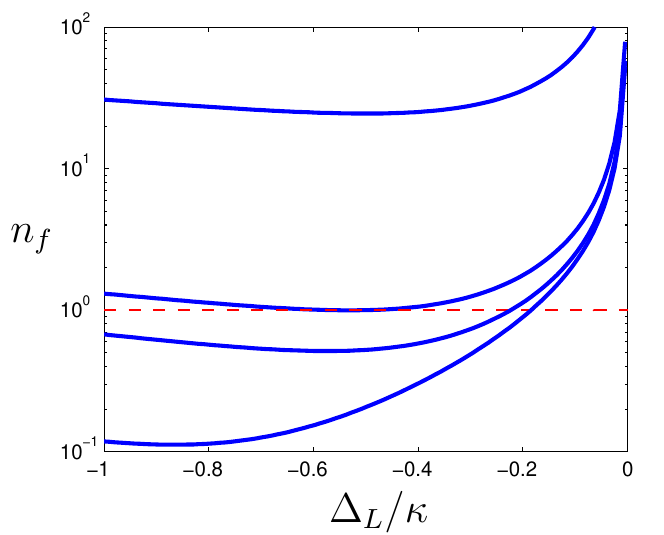}}
\caption{Final (steady state) average phonon number $n_f$ as a
  function of the normalized detuning $\Delta_L/\kappa$ for different
  values of the ratio $\kappa/\omega_m$, with
  $\gamma_m=n_p=0$.\label{fig:detuning}}
\end{figure}

The master Eq.~(\ref{eq:master}) generalizes the one obtained in
Ref.~\cite{WilsonRae07} by including thermal noise in the cavity
input. As shown below, this effect is significant for determining
the ultimate limit to which the mechanical resonator can be cooled for
ratios $\omega_p/\omega_m$ like the ones that characterize
electromechanical realizations. Equation (\ref{eq:master}) has as its
steady state a thermal state which defines the effective final
temperature to which the mechanical resonator is cooled. The
corresponding final occupancy reads [$\Gamma\equiv A_-(0)-A_+(0)$]:
\begin{equation}\label{eq:nfweak}
n_f= \frac{\gamma_m}{\Gamma + \gamma_m} n (\omega_m) +
\frac{\Gamma}{\Gamma+ \gamma_m} \left[ (2 \tilde{n}_f +1) n_p +
  \tilde{n}_f \right]  \,,
\end{equation}
where $\tilde{n}_f$ is the quantum backaction contribution derived in
Refs.~\cite{WilsonRae07,Marquardt07}, namely
\begin{equation}
\tilde{n}_f = - \frac{(\Delta_L + \omega_m )^2}{4 \Delta_L \omega_m}\,.
\end{equation}
We note that in the ``unshifted'' representation there is in addition
a coherent shift of the resonator's normal coordinate so that
\begin{equation}\label{eq:shift}
  \langle a_m^{\dag}a_m^{\vphantom{\dag}}\rangle= n_f + \eta^2 \left[
    \frac{P \kappa_{ex}}{\hbar \omega_L  (\Delta_L^2 + \kappa^2/4)}
  \right]^2\,. 
\end{equation} 
If we now consider the appreciable cooling limit $\Gamma\gg\gamma_m$
and minimize with respect to the detuning we obtain
\begin{equation}\label{eq:nfopt1}
\fl  \min \{n_f\} \approx
  \frac{\gamma_m\kappa\sqrt{\omega_m^2+\kappa^2/4}}{g_m^2\omega_m}\,
  n(\omega_m) \, +\, n_p\, +\, 
  \left(n_p+\frac{1}{2}\right)\!\left(\sqrt{1+\frac{\kappa^2}{4\omega_m^2}}
    -1\right)\,,
\end{equation}
for the optimal detuning
$\Delta_L^\mathrm{opt}=\sqrt{\omega_m^2+\kappa^2/4}$. The first term
corresponds to the ``linear cooling'' limited by thermal noise. In
turn, the second term shows that the final occupancy is necessarily
bounded by the equilibrium thermal occupancy of the cooler. Finally,
the last term for $n_p=0$ corresponds to the fundamental temperature
limit imposed by the quantum backaction which in the Doppler regime
$\kappa\ll\omega_m$ reduces to $\kappa/4\omega_m$, precluding ground
state cooling, while in the RSB regime it yields
$\kappa^2/16\omega_m^2$ corresponding to occupancies well below unity
\cite{WilsonRae07,Marquardt07}. We note that for a given
$\kappa/\omega_m<\sqrt{32}$ occupancies below unity are only
attainable within a finite detuning window
$|\Delta_L+3\omega_m|\leq\sqrt{8\omega_m^2 - \kappa^2/4}$, as
illustrated in Figure~\ref{fig:detuning}.

\subsection{Output power spectrum and temperature measurement}

The spectrum of the cavity output constitutes a crucial observable to
understand the backaction cooling. It allows to visualize the cooling
cycle as a frequency up-conversion process and for optomechanical
realizations it provides an efficient way to measure the final
temperature \cite{WilsonRae07}.  We focus on the latter for which
$n_p=0$ and consider the experimentally relevant case in which the
output is measured in the same modes in which the coherent laser drive
is fed \cite{Schliesser06}. To calculate its spectrum we apply the
standard input-output formalism \cite{Gardiner} and treat the
parameters $\eta|\alpha|$, $\eta$ perturbatively along the lines of
the previous Subsection. For this purpose it proves useful to consider
the quantum Langevin equation for $a_p$ associated to the Liouvillian
(\ref{eq:L}), namely
\begin{equation}\label{eq:langevin}
\fl  \dot{a}_p=\left(i\Delta_L-\frac{\kappa}{2}\right)a_p
  -i\eta\omega_m(a^{\vphantom{\dag}}_m +a^{\dag}_m)(a_p +
  \alpha)+\sqrt{\kappa_\mathrm{ex}} \delta
  a_\mathrm{in}(t)+\sqrt{\kappa-\kappa_\mathrm{ex}} b_\mathrm{in}(t)\,, 
\end{equation}
where $\delta a_\mathrm{in}$ and $b_\mathrm{in}$ correspond to the
vacuum noise associated, respectively, to the laser mode and to the
other cavity losses. Here we use the shifted representation
[cf.~(\ref{eq:hams})] and also take into account that Hamiltonian
(\ref{eq:hamun}) presupposes the standard time-dependent canonical
transformation that maps the coherent input state associated to the
laser into a classical field so that $a_\mathrm{in}(t)\to\delta
a_\mathrm{in}(t)+\langle a_\mathrm{in}(t)\rangle$. If we assume that
the term linear in $\eta$ is a specified function of time and that the
solution $a^{(0)}_p(t)$ for $\eta=0$ is known, Eq.~(\ref{eq:langevin})
can be formally integrated to obtain
\begin{equation}\label{eq:langint}
  \fl a_p(t) = a_p^{(0)}(t) -i \eta\omega_m \int_0^t \!\rmd \tau\ e^{(i \Delta_L
    - \kappa/2) (t-\tau)} \left[  a_m^\dag (\tau) + a_m^{\vphantom{\dag}}
    (\tau) \right] \left[a_p (\tau) + \alpha \right] 
  \,.
\end{equation}
This integral equation can be iterated to generate the following Dyson
series type result
\begin{eqnarray}\label{eq:Dyson}
  \fl a_p(t) = a_p^{(0)}(t) + \sum_{n=1}^{\infty} (- i\eta\omega_m)^n
  \int_0^t \!\rmd\tau_n \int_0^{\tau_n}  \!\rmd\tau_{n-1}  \ldots
  \int_0^{\tau_2}  \!\rmd\tau_1 \ e^{(i \Delta_L -\kappa/2) (t-\tau_1)}
  \left[ a_m^\dag (\tau_n) \right. \nonumber \\ \left.
    + a_m^{\vphantom{\dag}} (\tau_n) \right]
  \ldots  \left[ a_m^\dag (\tau_1) + a_m^{\vphantom{\dag}} (\tau_1)
  \right] \left[  a^{(0)}_p (\tau_1) + \alpha \right] 
  \,.
\end{eqnarray}

In turn, the output spectrum is given by
\begin{equation}
S(\omega)=\frac{1}{2\pi} \int_{-\infty}^{+\infty} \!\rmd \tau \
e^{-i (\omega - \omega_L ) \tau} \langle a^\dag_\mathrm{out} (t+\tau)
a^{\vphantom{\dag}}_\mathrm{out} (t)  \rangle_\mathrm{SS} \,,
\end{equation}
which in the shifted representation reads 
\begin{eqnarray}\label{eq:spect}
  \fl S(\omega)=\frac{1}{2\pi} \int_{-\infty}^{+\infty} \!\rmd \tau \
  e^{-i (\omega - \omega_L ) \tau} \langle \left[
    \sqrt{\kappa_\mathrm{ex}} \, a_p^\dag (t+\tau) + \delta
    a^\dag_\mathrm{in} (t+\tau)  + \sqrt{\kappa_\mathrm{ex}} \,
    \alpha^* + \langle a^{\vphantom{\dag}}_\mathrm{in} (0) \rangle^*
  \right] 
  \nonumber \\
  \times \left[ \sqrt{\kappa_\mathrm{ex}} \, a_p(t) + \delta
    a^{\vphantom{\dag}}_\mathrm{in} (t) + \sqrt{\kappa_\mathrm{ex}} \,
    \alpha + \langle a^{\vphantom{\dag}}_\mathrm{in} (0) \rangle
  \right] \rangle_\mathrm{SS}  \,.
\end{eqnarray}
We now seek the lowest non-trivial order in $\eta$. The output
$a_\mathrm{out}$ has a $c$ number part arising from the classical
drive and the cavity steady state amplitude and an operator part
corresponding to the fluctuations. Equation (\ref{eq:spect}) directly
implies that terms involving the $c$ numbers only contribute to the
``main line'' $\propto\delta(\omega-\omega_L)$. Furthermore, as in the
shifted representation the steady state of the electromagnetic modes
is the vacuum $|0\rangle$ it follows from Eqs.~(\ref{eq:Dyson}),
(\ref{eq:spect}) that the contribution of the cross term with the
operator part --- which vanishes if the cubic term in
Eq.~(\ref{eq:hams}) is omitted --- is at most higher order in $\eta$
and can be neglected relative to the contribution bilinear in the $c$
numbers. The latter corresponds to the classical reflection
coefficient that is straightforward to obtain considering the
near-resonant scattering into modes (real or fictitious) responsible
for the other losses $\kappa-\kappa_\mathrm{ex}$. Thus we arrive at
\begin{eqnarray}\label{eq:sapprox1}
\fl S(\omega)\approx \frac{P}{\hbar \omega_L} \left[ 1- \frac{
    (\kappa - \kappa_\mathrm{ex}) \kappa_\mathrm{ex}}{\Delta_L^2 +
    \kappa^2/4} \right] \delta (\omega - \omega_L) +
    \frac{\kappa_\mathrm{ex} g_m^2}{8\pi} 
  \nonumber \\ \times \int_{-\infty}^{+\infty} \!\rmd \tau \
  e^{-i (\omega - \omega_L ) \tau} \ \left< \int_0^{t+\tau}  \!\rmd\tau'_1 \
  e^{(-i \Delta_L -\kappa/2) (t+\tau-\tau'_1)} 
\left[ a_m^\dag (\tau'_1) + a_m^{\vphantom{\dag}} (\tau'_1) \right]
 \right. \nonumber \\ \qquad \times \left.
\int_0^{t}  \!\rmd\tau_1 \  e^{(i \Delta_L -\kappa/2) (t-\tau_1)} 
\left[ a_m^\dag (\tau_1) + a_m^{\vphantom{\dag}} (\tau_1) \right]
\right>_\mathrm{SS}\,, 
\end{eqnarray}
where we have used $a_p^{(0)} (t)|0\rangle =\delta
a_\mathrm{in}|0\rangle =0$ and the correction is higher order in
$\eta$ for all frequencies. To calculate the steady state two time
average in Eq.~(\ref{eq:sapprox1}) we adopt (as in the previous
Subsection) an interaction picture for the resonator mode
[i.e.~$a_m(t)\to e^{-i\omega_mt}a_m(t)$] and make the substitutions
$t+\tau -\tau'_1\to\tau'_1$, $t+\tau_1\to\tau_1$ in the time
integrals. In this representation the mechanical mode operators evolve
slowly compared with $1/\kappa$ so that in line with our derivation of
a Markovian master equation for the mechanical motion we can factor
them out of the time integrals whose upper limit can be extended to
infinity. This Markovian approximation yields
\begin{eqnarray}\label{eq:sapprox2}
  \fl \left< \int_0^{t+\tau}  \!\rmd\tau'_1 \
    e^{(-i \Delta_L -\kappa/2) (t+\tau-\tau'_1)} 
    \left[ a_m^\dag (\tau'_1) + a_m^{\vphantom{\dag}} (\tau'_1) \right]
    \int_0^{t}  \!\rmd\tau_1 \  e^{(i \Delta_L -\kappa/2) (t-\tau_1)} \right.
      \nonumber \\ \left. \vphantom{\int_0^{t}}\times \left[ a_m^\dag (\tau_1) +
      a_m^{\vphantom{\dag}} (\tau_1) \right] \right>_\mathrm{SS}
  \approx 2 \langle  \left[ G^*(\omega_m,0) e^{i \omega_m (t+\tau)} a_m^\dag
  (t+\tau)  + G^*(-\omega_m,0)\right. \nonumber \\ \left. \times
 e^{-i \omega_m (t+\tau)} a_m^{\vphantom{\dag}} (t+\tau) \right]
\left[ G(-\omega_m,0) e^{i \omega_m t} a_m^\dag (t) + G(\omega_m,0) e^{-i
    \omega_m t} a_m^{\vphantom{\dag}} (t) \right] \rangle_\mathrm{SS}
\,, 
\end{eqnarray}
where we have used Eq.~(\ref{eq:corr2}) with $n_p=0$. The two time
averages of the mechanical mode operators can now be calculated from
the master equation (\ref{eq:master}) using the quantum regression
theorem. The needed one time averages satisfy
\begin{eqnarray}\label{eq:aver}
  \fl \langle \dot{a}_m \rangle = - \left( i \Delta_m +
    \frac{\gamma_m+\Gamma}{2} \right) \langle a_m \rangle \,, \nonumber\\
  \fl \langle a^\dag_m a^{\vphantom{\dag}}_m \rangle_\mathrm{SS} = n_f  
  \,,\qquad 
  \langle  a^{\vphantom{\dag}}_m a^\dag_m \rangle_\mathrm{SS} = n_f +1   
  \,,\qquad 
  \langle  a^{\vphantom{\dag}}_m  a^{\vphantom{\dag}}_m \rangle_\mathrm{SS} =
  \langle  a^{\dag}_m a^\dag_m \rangle_\mathrm{SS} = 0\,.
\end{eqnarray}
Finally, a straightforward calculation leads from
Eqs.~(\ref{eq:sapprox1}), (\ref{eq:sapprox2}), (\ref{eq:aver}), and
(\ref{eq:corr0}) to the spectrum already given in
Ref.~\cite{WilsonRae07}, namely
\begin{eqnarray}\label{eq:spectrum}
  \fl S(\omega)\,\approx\, \frac{P}{\hbar \omega_L} \left[ 1- \frac{
      (\kappa - \kappa_\mathrm{ex}) \kappa_\mathrm{ex}}{\Delta_L^2 +
      \kappa^2/4} \right] \delta (\omega - \omega_L)+
  \frac{\kappa_\mathrm{ex}A_-n_f}{\kappa\pi}\,
  \frac{\frac{\gamma_\mathrm{eff}}{2}}{\left(\omega-\omega_L-\omega_m
      -\Delta_m \right)^2 +\frac{\gamma_\mathrm{eff}^2}{4}}  
  \nonumber\\ 
  +
  \frac{\kappa_\mathrm{ex}A_+\left(n_f+1\right)}{\kappa\pi}\,
  \frac{\frac{\gamma_\mathrm{eff}}{2}}{\left(\omega-\omega_L+\omega_m
      +\Delta_m \right)^2+ \frac{\gamma_\mathrm{eff}^2}{4}} \,,
\end{eqnarray}
where we have introduced $\gamma_\mathrm{eff}\equiv\gamma_m+\Gamma$
which is the total dissipation rate for the mechanical mode in the
presence of the drive that determines the linewidths of the motional
sidebands peaked at $\omega_L\pm\omega_m$. One should note that the
above corresponds to photon rate per unit frequency and that the
weight of the motional sidebands relative to the main line is of order
$\eta^2(2n_f+1)$ [$\eta^2\lesssim 10^{-8}$ for typical systems]. Here
unlike the case of atomic laser cooling \cite{Stenholm86,Leibfried03}
the mechanical dissipation induces an asymmetry in the weights of the
sidebands that allows to retrieve the final temperature directly from
the steady state. The ``blue'' sideband weighted by
$N_{-}=\frac{\kappa_\mathrm{ex}}{\kappa}A_{-}n_f$ corresponds to the
up-converted photons (anti-Stokes scattering) responsible for the
cooling while the ``red'' sideband weighted by
$N_{+}=\frac{\kappa_\mathrm{ex}}{\kappa}A_{+}(n_f+1)$ corresponds to
the down-converted photons (Stokes scattering) that result in heating.

It is interesting to note that the formalism of this Section does not
presuppose linearizing around the steady state and neglecting the cubic
term in Hamiltonian (\ref{eq:hams}) accordingly, but rather the
validity of such treatment emerges from a controlled procedure that
would allow to incorporate the necessary corrections if the intrinsic
non-linearity $\eta$ were larger. A straightforward self-consistency
criterion is to compare the steady state fluctuation of the cubic term
with that of the quadratic one. An heuristic estimate of their ratio
can be extracted from the total weight of the motional sidebands in
Eq.~(\ref{eq:spectrum}) which together with the analysis in
Subsec.~\ref{master} implies that the cubic term can be neglected
provided the conditions $\eta^2(2n_f+1)\omega^2\ll \kappa^2$ and
$|\alpha|^2\gg\eta^2$ are satisfied --- for the typical parameters in
cooling experiments these are always met.

\section{Linearized theory for coupled optomechanical
  modes}\label{nonpert}

\subsection{Small cavity linewidth limit  $\kappa\ll g_m$}

The treatment in the previous section is only applicable when the
cooling rate $A_-$ given by Eqs.~(\ref{eq:A2}), (\ref{eq:D2}),
(\ref{eq:A0}) is much smaller than $\kappa$ (note that the heating
rate $A_+$ is always bounded by $A_-$ for negative detuning). This
condition for arbitrary negative detuning results in
$g_m^2\ll\kappa^2$ so that it follows (as expected) that the motional
master equation is only warranted for small enough optomechanical
coupling. In the Doppler regime when this is violated the
aforementioned condition
$g_m\ll\max\{\sqrt{\omega_m\kappa},\,\omega_m\}$ underpinning the RWA
for the mechanical losses will also fail. In contrast in the resolved
sideband regime (RSB) relevant for ground state cooling
$\kappa\ll\omega_m$ implies that there is a wide parameter range of
interest in which Eq.~(\ref{eq:masteropt}) remains valid while
Eq.~(\ref{eq:master}) fails. Here we consider this RSB regime beyond
perturbative optomechanical coupling. As $g_m$ becomes comparable to
$\kappa$ it becomes necessary to follow the coupled dynamics of both
modes as described by Eq.~(\ref{eq:masteropt}) which for
$g_m>\kappa/2$ exhibits normal mode splitting. Though for a Gaussian
initial condition the approach to the steady state is always amenable
to a straightforward description, in the intermediate regime
$g_m\sim\kappa$ there will be no simple analog of
Eq.~(\ref{eq:master}) that allows to visualize the cooling process in
terms of phonon jumps. In turn, deep in the strong coupling regime but
away from the instability, i.e. for $\kappa\ll g_m\ll \omega_m$, the
dynamics can be described by two decoupled master equations for the
optomechanical normal modes analogous to Eq.~(\ref{eq:master}).

Within the latter parameter range it is permissible to start from a
Hamiltonian description including the optomechanical coupling and
treat the losses (as described by $\gamma_m,\,\kappa$) perturbatively.
We focus on the ``resonant case'' $-\Delta_L=\omega_m$ --- which in
the RSB regime can be shown to be optimal for minimizing the final
occupancy --- and consider the canonical transformation that
diagonalizes Hamiltonian (\ref{eq:hams}) for $\eta=0$ with $g_m\neq0$
[after performing the convenient rotation
$a_p\to(\alpha/|\alpha|)a_p$]. The latter is given by
\begin{eqnarray}\label{eq:canonic}
  \fl
  a_{m/p}=\frac{1}{2\sqrt{2}}\left[\left(\sqrt{\frac{\omega_m}{\omega_+}}
      +\sqrt{\frac{\omega_+}{\omega_m}}\right)a_+ \pm
    \left(\sqrt{\frac{\omega_m}{\omega_-}} 
      +\sqrt{\frac{\omega_-}{\omega_m}}\right)a_- \right.\nonumber\\\left.
    +\left(\sqrt{\frac{\omega_m}{\omega_+}}
      -\sqrt{\frac{\omega_+}{\omega_m}}\right)a_+^{\dag} \pm
    \left(\sqrt{\frac{\omega_m}{\omega_-}} 
      -\sqrt{\frac{\omega_-}{\omega_m}}\right)a_-^{\dag}\right]\,, 
\end{eqnarray}
where the eigenfrequencies of the normal modes read
\begin{equation}\label{eq:omegapm}
\omega_\pm=\omega_m(1\pm g_m/\omega_m)^{1/2}\,.
\end{equation} 
If we now consider the expansion of the above in the small parameter
$g_m/\omega_m$ the zeroth order of Eq.~(\ref{eq:omegapm}) yields a
splitting given by $g_m$ while Eq.~(\ref{eq:canonic}) reduces to the
transformation that diagonalizes the rotating wave part of Hamiltonian
(\ref{eq:hams}) --- that results from neglecting the terms that
involve $a_ma_p$, $a_m^{\dag}a_p^{\dag}$.  The latter transformation
does not mix the annihilation and creation operators. For small but
finite $g_m/\omega_m$ there will be small admixtures that for the
purpose of analyzing the cooling dynamics will only be relevant
insofar as they give rise to \emph{qualitatively new terms} in the
dissipative part of the Liouvillian --- otherwise they can be shown to
result in contributions of relative order $(g_m/\omega_m)^2$ for all
values of the other parameters. Hence we have
\begin{equation}\label{eq:canonicapr}
a_{m/p}\approx \frac{1}{\sqrt{2}}a_+ \pm \frac{1}{\sqrt{2}}a_-
-\frac{g_m}{4\sqrt{2}\omega_m}a_+^{\dag} 
\pm\frac{g_m}{4\sqrt{2}\omega_m}a_-^{\dag}\,.
\end{equation}

To proceed we: (i) apply the transformation given by
Eq.~(\ref{eq:canonic}) to the master equation (\ref{eq:masteropt}),
(ii) transform the result to an interaction picture with respect to
the (now diagonal) Hamiltonian, (iii) neglect all the resulting fast
rotating terms which are $\propto e^{\pm 2 i\omega_m t}$ or $\propto
e^{\pm i g_m t}$ (up to corrections higher order in $g_m/\omega_m$)
and (iv) expand to lowest order in $g_m/\omega_m$ following the
aforementioned ``qualitative'' criterion. Naturally (iii) relies on
the small cavity linewidth condition $\kappa\ll g_m$. Thus we obtain
\begin{eqnarray}
  \fl \dot{\rho}=
  \sum_{\xi=\pm}\left[\frac{\gamma_m}{4}n(\omega_m)
    +\frac{\kappa}{4}n(\omega_p) + \frac{\kappa
      g_m^2}{64\,\omega_m^2}\right]
  \!\left(2a_\xi^{\dag}\rho
    a_\xi-a_\xi a_\xi^{\dag}\rho-\rho a_\xi a_\xi^{\dag}\right) \nonumber \\
  + \left\{\frac{\gamma_m}{4}[n(\omega_m)+1]
    +\frac{\kappa}{4}[n(\omega_p)+1]\right\}\!  
  \left(2a_\xi\rho
    a_\xi^{\dag}-a_\xi^{\dag} a_\xi\rho-\rho a_\xi^{\dag} a_\xi\right)
  \,.\label{eq:masterstrong} 
\end{eqnarray}
The only corrections in $(g_m/\omega_m)^2$ appear in the first term
and correspond to the ``high power limit'' of the heating induced by
the quantum backaction of the cavity. Analogous heating terms $\propto
\gamma_m$ are neglected given that they are comparable to corrections
to the RWA treatment of the mechanical dissipation. It follows from
Eq.~(\ref{eq:masterstrong}) that the losses do not couple the normal
modes (annihilation operators $a_\pm$) so that the steady state is
given by the tensor product of thermal states for each of them
characterized by average occupancies
\begin{equation}\label{eq:avstrong}
  \langle a_\pm^{\dag}
  a_\pm^{\vphantom{\dag}}\rangle_\mathrm{SS}=\frac{\gamma_m
    n(\omega_m)+\kappa n(\omega_p)+\frac{\kappa
      g_m^2}{16\,\omega_m^2}}{\gamma_m +\kappa}
\end{equation}
(where we neglect higher order terms in $g_m/\kappa$) to which the
average phonon number converges with a cooling rate
$(\kappa+\gamma_m)/2$. Hence Eqs.~(\ref{eq:canonicapr}),
(\ref{eq:avstrong}) finally yield
\begin{eqnarray}\label{eq:nfstrong}
  \fl  n_f =\frac{\gamma_m
    n(\omega_m)+\kappa n(\omega_p)+\frac{\kappa
      g_m^2}{16\,\omega_m^2}}{\gamma_m +\kappa}+
  \frac{g_m^2}{16\,\omega_m^2}\nonumber\\
  \approx \frac{\gamma_m
    n(\omega_m)}{\kappa} + n_p +\frac{g_m^2}{8\omega_m^2}
\end{eqnarray}
where the corrections are higher order in the small parameters
$(g_m/\omega_m)^2$, $\gamma_m/\kappa$. The first term corresponds to
the heating associated to the mechanical dissipation and can be
identified with the corresponding term in Eq.~(\ref{eq:nfweak})
showing a saturation of the usual linear cooling law. Similarly, the
second and third terms can be identified with the corresponding
contributions in Eq.~(\ref{eq:nfweak}) arising from the thermal noise
in the cavity and the quantum backaction. Thus ground state cooling
requires $k_B T/\hbar Q_m\ll \kappa \ll\omega_m$ and $\omega_p<k_B
T$. We note that Eqs.~(\ref{eq:canonicapr}), (\ref{eq:avstrong}) imply
that $\langle a_m^2\rangle_\mathrm{SS}=0$ so that the reduced state
for the motion is also thermal in the small $\kappa$ limit.

If we now compare Eqs.~(\ref{eq:nfopt1}) and (\ref{eq:nfstrong}) and
consider minimizing them with respect to $g_m$ within their respective
ranges of validity, heuristic considerations imply that the following
formula is expected to always constitute a lower bound for the final
mechanical occupancy optimized with respect to the parameters of the
drive
\begin{equation}\label{eq:nfopt2}
  n_\mathrm{TL}= \frac{\gamma_m
    n(\omega_m)}{\kappa} + n_p + \frac{1}{2}
  \left(\sqrt{1+\frac{\kappa^2}{4\omega_m^2}}
    -1\right)\,.
\end{equation}

\subsection{Final occupancy for arbitrary ratio $g_m/\kappa$ and
  optimal parameters}

The approximate expressions (\ref{eq:nfweak}), (\ref{eq:nfstrong}) for
the final mechanical occupancy that we have derived in the limits
$g_m\ll\kappa$ and $g_m\gg\kappa$ provide a basic understanding of the
requirements for ground state cooling and the expected order of
magnitude for the optimum. Notwithstanding they have the drawback that
they fail to settle which is the optimal input power as minimization
with respect to $g_m$ shifts this variable away from the domain where
they are valid. In addition given the experimental progress towards
achieving ultra-cold states in these systems it is clear that precise
quantitative predictions of the steady state for a given input are
highly desirable. To this effect we complement the above analysis by
deriving an analytical expression for $n_f$ valid for arbitrary values
of the ratio $g_m/\kappa$. The optomechanical master equation
(\ref{eq:masteropt}) directly implies that when the cubic
non-linearity in Hamiltonian (\ref{eq:hams}) is neglected the time
evolution for the ten independent second order moments that determine
the covariance matrix is given by a linear system of ODEs. This is
determined by
\begin{eqnarray}\label{eq:momexact}
  \fl \frac{\rmd }{\rmd t} \langle a_m^\dag a_m^{\vphantom{\dag}}
  \rangle = - \gamma_m \langle a_m^\dag a_m^{\vphantom{\dag}} \rangle
  + \gamma_m n (\omega_m) - i \frac{g_m}{2} \langle  \left( a_p^\dag +
    a_p^{\vphantom{\dag}} \right) \left(  a_m^\dag -
    a_m^{\vphantom{\dag}} \right) \rangle  \,,\\ 
  \fl \frac{\rmd }{\rmd t} \langle  a_p^{\vphantom{\dag}}
  a_m^{\vphantom{\dag}} \rangle = - \left( \frac{\kappa}{2} +
    \frac{\gamma_m}{2} - i \Delta_L + i \omega_m \right) \langle
  a_p^{\vphantom{\dag}} a_m^{\vphantom{\dag}} \rangle
\nonumber \\ 
 - i \frac{g_m}{2}
  \left( 1 + \langle a_m^\dag a_m^{\vphantom{\dag}} \rangle +
    \langle a_m^2 \rangle + \langle a_p^\dag a_p^{\vphantom{\dag}}
    \rangle  + \langle a_p^2 \rangle  \right)  \,,\\ 
  \fl \frac{\rmd }{\rmd t} \langle  a_p^{\vphantom{\dag}}
  a_m^\dag \rangle = - \left( \frac{\kappa}{2} +
    \frac{\gamma_m}{2} - i \Delta_L - i \omega_m \right) \langle
  a_p^{\vphantom{\dag}} a_m^\dag \rangle 
\nonumber \\ - i \frac{g_m}{2}
  \left( \langle a_m^\dag a_m^{\vphantom{\dag}} \rangle +  \langle
    a_m^2 \rangle - \langle a_p^\dag a_p^{\vphantom{\dag}}
    \rangle - \langle a_p^2 \rangle \right) \,, \\
  \fl \frac{\rmd }{\rmd t} \langle a_p^\dag a_p^{\vphantom{\dag}}
  \rangle = - \kappa \langle a_p^\dag a_p^{\vphantom{\dag}} \rangle -
  i \frac{g_m}{2} \langle  \left( a_p^\dag -
    a_p^{\vphantom{\dag}} \right) \left(  a_m^\dag +
    a_m^{\vphantom{\dag}} \right) \rangle + \kappa n (\omega_p) \,, \\
  \fl \frac{\rmd }{\rmd t} \langle a_m^2 \rangle = - \left( \gamma_m +
  2 i \omega_m \right) \langle a_m^2 \rangle -i g_m
\langle  \left( a_p^\dag + a_p^{\vphantom{\dag}} \right)
a_m^{\vphantom{\dag}} \rangle \,, \\ 
\fl \frac{\rmd }{\rmd t} \langle a_p^2 \rangle = - \left( \kappa - 2 i
  \Delta_L \right) \langle a_p^2 \rangle - i g_m 
\langle  \left( a_m^\dag + a_m^{\vphantom{\dag}} \right)
a_p^{\vphantom{\dag}} \rangle \,,
\end{eqnarray}
and their Hermitian conjugates. The exact solution of the linear
system of equations that results for the steady state covariance
matrix then yields an analytical formula for the final occupancy of
the mechanical resonator that is a sum of three independent
contributions
\begin{equation}\label{eq:nfexact1}
n_f^{\vphantom{(0)}}= n_f^{(m)} + n_f^{(p)} + n_f^{(0)} \,.
\end{equation}
Here $n_f^{(m)}$ and $n_f^{(p)}$ arise, respectively, from the
mechanical dissipation and the thermal noise in the cavity input, and
$n_f^{(0)}$ corresponds to the heating induced by the quantum
backaction exerted by the cavity. The expressions for these different
contributions are rather unwieldy for arbitrary mechanical dissipation
$\gamma_m$, but it is simple to realize that for appreciable cooling
$n_f\ll n(\omega_m)$ to be possible the conditions $\gamma_m\ll
\kappa,\,g_m,\,\omega_m$ are needed. Hence in this regime of interest
a good approximation for $n_f$ is obtained if one takes $\gamma_m\to0$
keeping $\gamma_m n(\omega_m)$ finite. In this limit we obtain
\begin{eqnarray}\label{eq:nfexact2}
\fl n_f^{(0)} = - \frac{1}{4 \omega_m \Delta_L} \left[\left(
  \omega_m+ \Delta_L \right)^2 + \frac{\kappa^2}{4} \right] +
\frac{g_m^2\,R}{8 \omega_m} \left( \Delta_L^2 +
  \frac{\kappa^2}{4}\right)   \,,\nonumber\\ 
\fl n_f^{(m)} = -\frac{\gamma_n n(\omega_m)\,R}{\kappa \Delta_L g_m^2
  \omega_m^2} \left\{ \frac{\Delta_L^2 g_m^4}{4} \left( \Delta_L +
    \frac{\kappa^2}{4} -4 \omega_m^2 \right) + \omega_m^2 \left(
    \Delta_L^2 + \frac{\kappa^2}{4} \right) \left[
    \left( \omega_m- \Delta_L \right)^2 + \frac{\kappa^2}{4} \right]
\right. \nonumber \\ \left. \times
   \left[ \left( \omega_m+ \Delta_L \right)^2 + \frac{\kappa^2}{4}
   \right] + \Delta_L g_m^2 \omega_m \left(\omega_m^4 + \Delta_L^4 +
   \Delta_L^2 \frac{\kappa^2}{2} - 3 \omega_m^2 \Delta_L^2 +
   \frac{\kappa^4}{16} 
\right. \right. \nonumber \\ \left. \left. 
+ \omega_m^2 \frac{\kappa^2}{4}  \right) \right\} \,,\nonumber\\
\fl n_f^{(p)} = - \frac{n_p\,R}{2 \Delta_L\omega_m} \left[
  \Delta_L \left( \omega_m \Delta_L^3 + \omega_m^3 \Delta_L + g_m^2
    \Delta_L^2/2 + \omega_m^2 g_m^2 \right) + \left( 2 \omega_m
    \Delta_L^2 +  g_m^2 \Delta_L/2
\right.  \vphantom{\frac{\kappa^2}{4}}\right. \nonumber \\ \left. \left. 
+ \omega_m^3 \right) \frac{\kappa^2}{4} + \omega_m \frac{\kappa^4}{16}
\right] \,,
\end{eqnarray}
with 
\begin{equation}
R = \frac{1}{ \left[ \Delta_L \left( \omega_m \Delta_L + g_m^2 \right) +
 \omega_m  \frac{\kappa^2}{4} \right] } \,.
\end{equation}
It is straightforward to show that in the limits $g_m\ll\kappa$ with
$\Delta_L<0$, and $\kappa\ll g_m\ll \omega_m$ with
$\Delta_L=-\omega_m$ we recover, respectively, the approximate
expressions (\ref{eq:nfweak}), (\ref{eq:nfstrong}) [to lowest order in
$\gamma_m/\Gamma$].

\begin{figure}
\centerline{\includegraphics[width=\linewidth]{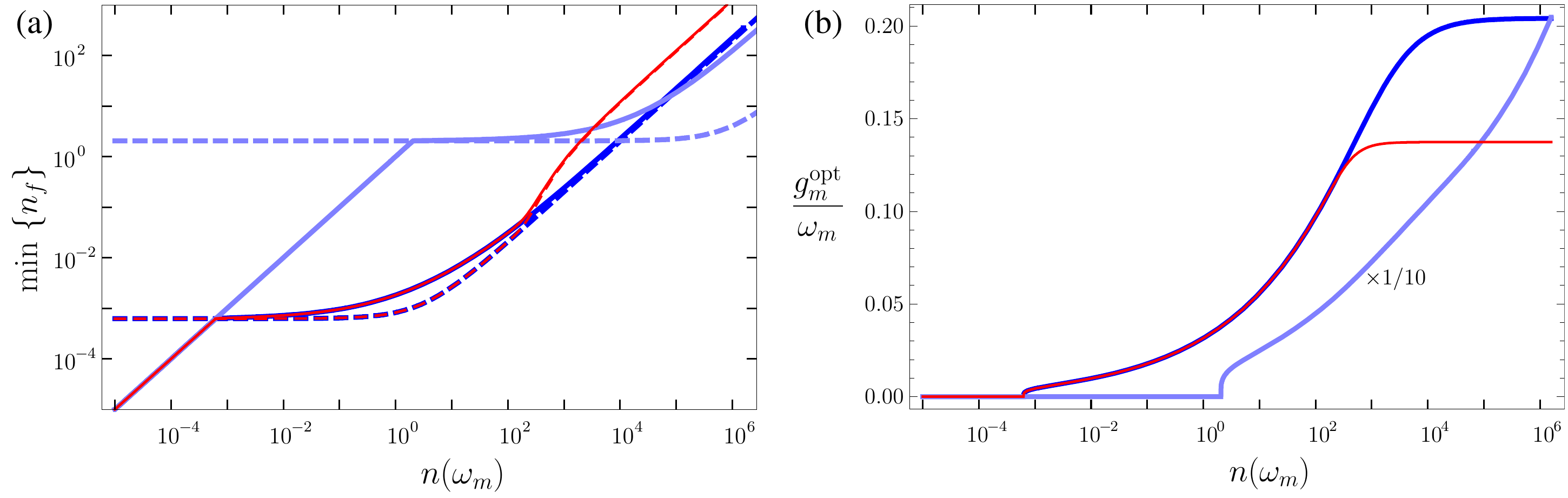}}
\caption{(a) Final (steady state) average phonon number minimized with
  respect to $g_m$ and $\Delta_L$ [$\min\{n_f\}$] as a function of the
  thermal mechanical occupancy $n(\omega_m)$ [solid lines]. 
  The lower bound $n_{\mathrm{TL}}$ given by Eq.~(\ref{eq:nfopt2}) is
  shown for comparison [dashed
  lines]. 
(b) Optimal normalized optomechanical coupling
$g_m^\mathrm{opt}/\omega_m$ for which the minimum mechanical
occupancies shown in (a) are attained.
In both (a) and (b), $Q_m=5\times10^{4}$, the light blue lines
correspond to $n_p=0$ and $\kappa/\omega_m=10$, the dark blue lines to 
$n_p=0$ and $\kappa/\omega_m=0.1$, and the red lines to
$\kappa/\omega_m=0.1$ and a finite $n_p$ determined by
$\omega_p/\omega_m=10^{3}$. 
\label{fig:opt}}
\end{figure}

\begin{figure}
\centerline{\includegraphics[width=\linewidth]{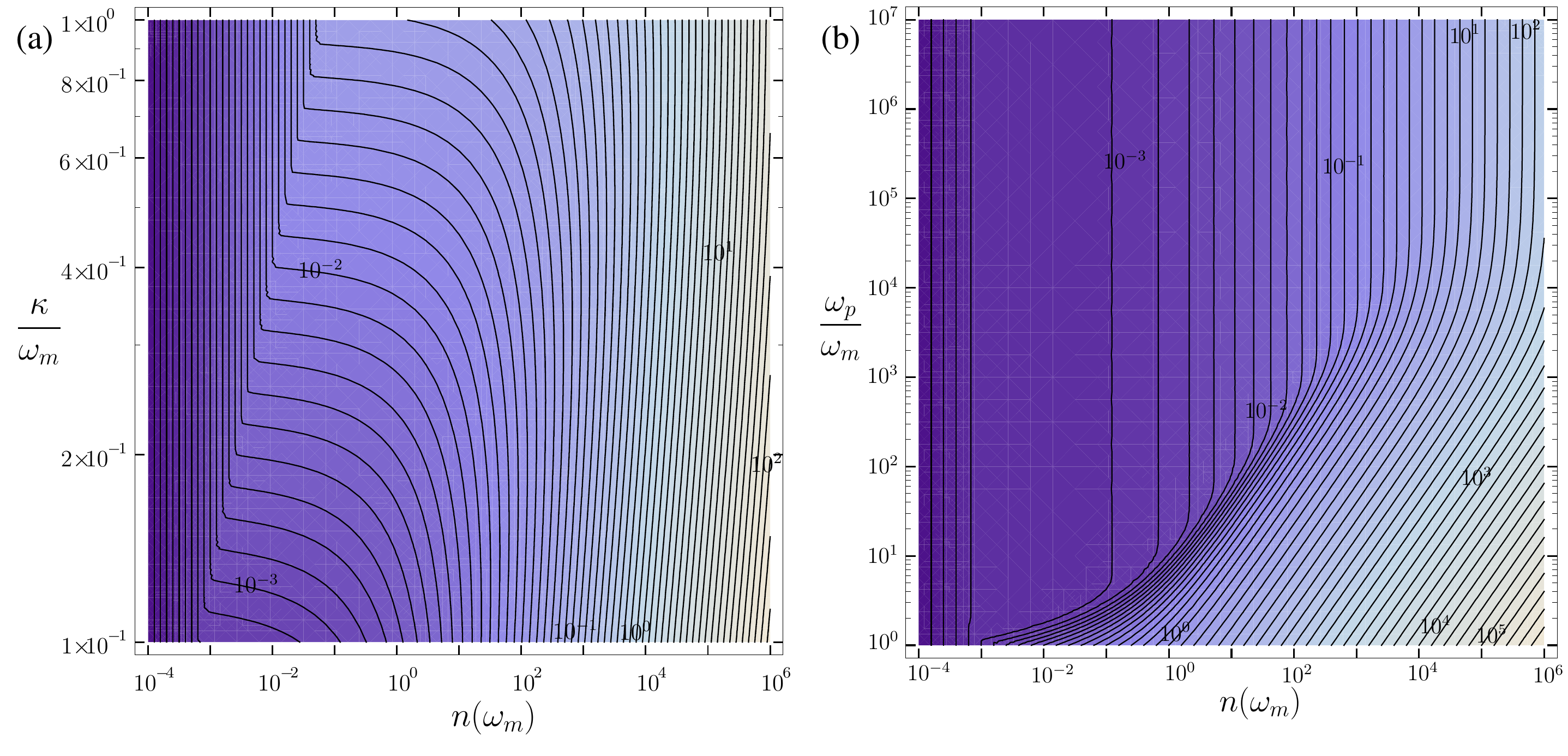}}
\caption{(a) Final (steady state) average phonon number minimized with
  respect to $g_m$ and $\Delta_L$ [$\min\{n_f\}$] as a function of the
  thermal mechanical occupancy $n(\omega_m)$ and the ratio
  $\kappa/\omega_m$, for $Q_m=5\times10^{4}$ and $n_p=0$. (b)
  \emph{Idem} as a function of the
  thermal mechanical occupancy $n(\omega_m)$ and the ratio
  $\omega_p/\omega_m$, for $Q_m=5\times10^{4}$ and $\kappa/\omega_m=0.1$.
  \label{fig:nf}}
\end{figure}

The analytical formulas Eq.~(\ref{eq:nfexact1}), (\ref{eq:nfexact2})
provide a quantitative basis for a precise analysis of the final
temperatures that can be attained for arbitrary values of
$g_m/\kappa$. In particular they can be readily minimized numerically
with respect to the \emph{drive's detuning and input power} subject to
the constraint imposed by the stability condition $R^{-1}>0$ that
emerges from the Routh-Hurwitz criteria. This is the natural
optimization problem that is posed by an experimental realization in
which the natural linewidth of the cavity is fixed or hard to modify
(admittedly sweeping this parameter within some range should be
straightforward in electromechanical setups). We have performed this
optimization using instead the exact expression that follows from
Eq.~(\ref{eq:momexact}), i.e.~without the small $\gamma_m$
approximation so that the transition to the trivial regime
$\min\{n_f\}=n(\omega_m)$ is captured. Figure~\ref{fig:opt}(a)
compares this optimum (solid lines) as a function of the thermal
equilibrium occupancy $n(\omega_m)$ with the lower bound
$n_\mathrm{TL}$ (dashed lines) furnished by Eq.~(\ref{eq:nfopt2}) for
representative values of the other parameters. The blue lines
correspond to $n_p=0$ (i.e. $\omega_p/\omega_m\to\infty$) for values
of $\kappa/\omega_m$ in the Doppler regime (light blue) and in the RSB
regime (dark blue). The red lines show the latter with
$\omega_p/\omega_m=10^{3}$ instead. There are three distinct regimes:
(i) $n(\omega_m)<n_\mathrm{TL}$ so that the optomechanical coupling
only raises the temperature and $g_m^\mathrm{opt}=0$, (ii) the minimum
occupancy is determined by the fundamental quantum backaction limit
and is thus independent of the ambient temperature [i.e.~constant as a
function of $n(\omega_m)$], and (iii) the minimum occupancy is
determined by classical noise and is thus linear in the ambient
temperature. In this latter regime for finite $\omega_p/\omega_m$
there is a noticeable transition at which the minimum deviates from
the $n_p=0$ result as the cavity becomes thermally activated. One can
also note that in this linear regime $n_\mathrm{TL}$ is only reachable
in the RSB limit. In turn Fig.~\ref{fig:opt}(b) shows the
corresponding optimal values for $g_m$. The saturation at high
temperatures just reflects the fact that
$(n_f^{m}+n_f^{p})/n(\omega_m)$ becomes temperature independent in
that limit and attains its minimum at a finite $g_m$
[c.f.~(\ref{eq:nfexact2})]. Finally, Fig.~\ref{fig:nf}(a) and (b)
illustrate, respectively, the dependence of the optimum with the
ratios $\kappa/\omega_m$ and $\omega_p/\omega_m$ as the ambient
temperature is varied in the RSB regime. The fundamental quantum
backaction limit results in a distinct shoulder. In
Fig.~\ref{fig:nf}(a) its lower edge traces a line with a slope
consistent with the quadratic dependence on
$\kappa/\omega_m$. It is clear from Fig.~\ref{fig:nf}(b) that for
ratios $\omega_p/\omega_m\lesssim10^{4}$ and
$n(\omega_m)\gtrsim10^{3}$ (parameters that are relevant for
electromechanical setups \cite{Brown07}) thermal noise in the cavity
needs to be taken into account.

\section{Conclusion}

In summary, we have analyzed the cavity-assisted backaction cooling of
a mechanical resonator in the quantum regime by deriving effective
master equations for the relevant degrees of freedom both in the
perturbative and strong coupling regimes. These provide a description
of the cooling dynamics that allows to establish a simple lower bound
for the final occupancy [Eq.~(\ref{eq:nfopt2})] that can only be
reached in the RSB regime. This bound implies that ground state
cooling is only possible when the cavity linewidth is much larger than
the heating rate induced by the mechanical dissipation but much
smaller than the mechanical oscillation frequency, and the equilibrium
thermal occupancy of the cavity is well below unity. In addition we
give an analytical expression for the final occupancy valid in all
regimes of interest that allows for a straightforward optimization of
the parameters of the drive. Finally we analyze the dependence of this
optimum on the ambient temperature, the cavity linewidth, and the
ratio of the cavity's frequency to the mechanical frequency.

\ack 

NN, TJK, and WZ acknowledge funding via the Nanosystems
Initiative Munich.

\section*{References}
\bibliographystyle{unsrt}

\end{document}